**Title of the Manuscript:**

**Has the Paris Agreement Shaped Emission Trends? A Panel VECM Analysis of Energy, Growth, and $CO_2$ in 106 Middle-Income Countries**

**List of Authors and Information:**

1. Tuhin G M Al Mamun

    Department of Economics

    Hannam University, South Korea

    Email-20224130@gm.hannam.ac.kr

2. Ehsanullah

    Department of Economics

    Hannam University, South Korea

    Email- ehsan3171@gmail.com

3. Md Sharif Hassan

    School of Accounting and Finance,

    Taylor's University, Malaysia

    Email: mdsharifhassan27@gmail.com; MdSharif.Hassan@taylors.edu.my

    ORCID: 0000-0001-8418-7578

4. Mohammad Bin Amin*
    Doctoral School of Management and Business

    Faculty of Economics and Business,

    University of Debrecen, Böszörményi út 138; Post Code: 4032,

    Debrecen, Hungary.

    E-mail: binamindu@gmail.com

    ORCID: 0000-0002-9184-4828

    &

    Department of Business Administration,

    Faculty of Business Studies,

Bangladesh Army University of Science and Technology,

Saidpur, Nilphamari-5310, Rangpur, Bangladesh

5. Judit Oláh
   Faculty of Economics and Business,

   University of Debrecen,

   Böszörményi út 138; 4032 Debrecen, Hungary,

   Scopus Author ID: 56016286600

   E-mail: olah.judit@econ.unideb.hu

   https://orcid.org/0000-0003-2247-1711

   &

   Doctoral School of Management and Business Administration,

   John von Neumann University

   6000 Kecskemét, Hungary,

   &

   Department of Trade and Finance, Faculty of Economics and Management,

   Czech University of Life Sciences Prague, Czech Republic

*Correspondence Author: Mohammad Bin Amin (binamindu@gmail.com)

Funding: None.

Acknowledgments: This research was supported by the "*University of Debrecen Program for Scientific Publication*".

**Abstract**

Rising $CO_2$ emissions remain a critical global challenge, particularly in middle-income countries where economic growth drives environmental degradation. This study examines the long-run and short-run relationships between $CO_2$ emissions, energy use, GDP per capita, and population across 106 middle-income countries from 1980 to 2023. Using a Panel Vector Error Correction Model (VECM), we assess the impact of the Paris Agreement (2015) on emissions while conducting cointegration tests to confirm long-run equilibrium relationships.

The findings reveal a strong long-run relationship among the variables, with energy use as the dominant driver of emissions, while GDP per capita has a moderate impact. However, the Paris Agreement has not significantly altered emissions trends in middle-income economies. Granger causality tests indicate that energy use strongly causes emissions, but GDP per capita and population do not exhibit significant short-run causal effects. Variance decomposition confirms that energy shocks have the most persistent effects, and impulse response functions (IRFs) show emissions trajectories are primarily shaped by economic activity rather than climate agreements. Robustness checks, including autocorrelation tests, polynomial root stability, and Yamagata-Pesaran slope homogeneity tests, validate model consistency.

These results suggest that while global agreements set emissions reduction goals, their effectiveness remains limited without stronger national climate policies, sectoral energy reforms, and financial incentives for clean energy adoption to ensure sustainable economic growth.

**Introduction**

Climate change has become one of the most pressing global challenges, with carbon dioxide ($CO_2$) emissions being a primary driver of global warming. The relationship between economic growth, energy consumption, and $CO_2$ emissions has long been debated in environmental economics, particularly as policymakers strive to balance economic development with sustainability. Over the past few decades, rapid industrialization and population growth have led to a surge in global energy demand, contributing to higher emissions and environmental degradation. Numerous international agreements have emerged to reduce emissions without disrupting economic stability. The Paris Agreement (2015) stands as the most extensive worldwide initiative to both decrease emissions and promote sustainable energy adoption. The true impact of international agreements with ambitious climate goals on changing emissions paths continues to be ambiguous.

Policymakers need to determine if international climate treaties such as the Paris Agreement significantly changed the dynamics between carbon dioxide emissions and energy consumption alongside GDP per capita and population growth. Numerous countries have pledged to cut emissions but the empirical data shows inconsistent results regarding their actual impact on measurable outcomes. To develop successful policies that support sustainable development while maintaining economic growth requires a deep comprehension of these dynamics. The Environmental Kuznets Curve (EKC) hypothesis has been researched extensively and proposes that economic growth leads to increased emissions before they begin to decrease as economies adopt cleaner technologies. Current studies on economic activity and emissions relationships do not consider structural policy changes such as the Paris Agreement because they assume static connections. Current econometric models fail to analyze how climate policies establish permanent changes in emission patterns which creates a theoretical void in evaluating policy-driven structural changes.

To address this gap, this research explores the long-term and short-term relationships among $CO_2$ emissions, energy consumption, GDP per capita, and population growth by including the Paris Agreement as a structural break in a Panel Vector Error Correction Model (VECM). This study uses data from 106 middle-income countries between 1980 and 2023 to determine how effective the Paris Agreement has been in influencing emissions while assessing whether economic and demographic factors independently drive emissions regardless of policy measures. The substantial portion of global emissions generated by middle-income countries necessitates an examination of how international climate agreements affect their emission patterns. Middle-income nations struggle to balance economic growth with environmental commitments even though high-income countries have made substantial progress toward decarbonization, which makes these countries key targets for detailed empirical analysis.

To explore these issues, the study seeks to answer the following research questions:

1. What is the long-run relationship between $CO_2$ emissions, energy use, GDP per capita, and population across a panel of middle-income countries?
2. Has the Paris Agreement significantly altered the relationship between these variables?
3. How do short-run adjustments in emissions and economic factors respond to climate policy changes?

By integrating the Paris Agreement into a VECM framework, this study contributes to the literature by explicitly testing for structural breaks in emissions trends, examining both short-term fluctuations and long-term equilibrium relationships. Furthermore, a homogeneity test is conducted to assess whether the determinants of emissions vary across countries, providing a nuanced understanding of emission dynamics in middle-income economies.

The remainder of this paper is structured as follows. The next section reviews the relevant literature on $CO_2$ emissions, economic growth, energy use, and climate agreements, providing a theoretical foundation for understanding emissions trends. The methodological framework follows, outlining the data sources, model specification, and econometric techniques employed in the study.

The empirical findings then presented including cointegration tests followed by VECM estimations and robustness examinations. The study findings are discussed through the lens of the research questions to emphasize the resulting implications for policymakers. Finally, The conclusion of the paper summarizes major findings and discusses literature contributions while acknowledging study constraints and proposing future research directions.

**Literature Review**

Numerous studies investigate how $CO\_2$ emissions interact with economic growth and energy consumption. The researchers evaluate how international climate agreements affect environmental conditions. This section reviews existing research in three key areas: This section examines three research areas including the EKC hypothesis which describes emissions changes relative to economic growth, the connection between energy consumption and emissions levels, and the success of climate agreements like the Paris Agreement.

Environmental economics research often examines the Environmental Kuznets Curve (EKC) hypothesis which serves as a fundamental concept in the field. Environmental degradation escalates alongside economic growth but eventually decreases once countries reach a specific level of income. The initial stages of economic development boost $CO\_2$ emissions because industrialization leads to increased energy consumption. Emissions start to decline when countries transition to cleaner technologies and implement stricter environmental regulations (Grossman & Krueger, 1995; Stern, 2004).

Early empirical research supported the EKC hypothesis. Subsequent research indicated country-specific inconsistencies emerged from differences in policy frameworks as well as institutional quality and energy sources (Dinda, 2004; Narayan & Narayan, 2010). According to Stern (2017) some researchers dispute the EKC by claiming that emissions reductions cannot happen solely through economic growth. The crucial role of renewable energy adoption, technological advancements, and environmental regulations demonstrates that they are key factors according to Shahbaz et al. (2013).

Research indicates that apparent emissions reductions in wealthy countries may not accurately reflect real progress. Certain industries transfer their high-emission production processes to developing countries which moves emissions geographically instead of reducing them worldwide (Peters et al., 2011).Global carbon emissions require policy solutions due to their intricate emission patterns.

The analysis of economic growth's impact on CO2 emissions is essential particularly for middle income countries. Different countries experience different levels of industrialization and economic transition processes. Emissions trajectories for these countries will depend on their economic growth patterns alongside their energy consumption habits and policy decisions. Several studies have demonstrated that energy consumption directly impacts carbon emissions levels. Research findings show fossil fuel usage as a leading cause of emissions (Sadorsky, 2009, Shahbaz et al. ,2013).

Data demonstrates that increased energy use from non-renewable sources results in higher $CO_2$ emissions. Energy efficiency measures and the shift to renewable energy play a vital role in climate mitigation efforts according to Stern (2011).

Multiple research efforts investigate how energy efficiency along with renewable energy sources contribute to emissions reduction. Nations that incorporate more renewable energy sources within their energy systems show reduced rates of emissions growth.The transition speed differs among regions because of their unique economic frameworks and varying technological capabilities and policy standards (Apergis & Payne, 2010; Wang et al., 2016).

Certain researchers assert that renewable energy policies must be accompanied by other initiatives to succeed. According to Popp (2012), a combination of carbon pricing along with clean energy technology investments is essential. Uncertainty exists regarding how well energy transition policies work in middle-income countries. It is essential to understand the impact of energy consumption patterns on emission trends within these economies.

The impact of international climate agreements such as the Kyoto Protocol and the Paris Agreement is widely debated. Some studies suggest that these climate agreements have contributed to emissions reductions through various ways.Such as policy coordination, emissions targets and financial mechanisms (Aichele & Felbermayr, 2013).Others argue that weak enforcement mechanisms limit the effectiveness of the climate agreements.In Accordamce with the study of Böhringer et al.,( 2012), Emissions reductions often result from economic slowdowns rather than policy commitments .

The Kyoto Protocol introduced legally binding emissions targets for developed countries but faced criticism. Major emitters such as the United States did not participate and on the other hand, developing countries had no binding commitments (Victor, 2011).The Paris Agreement replaced Kyoto agreement with voluntary national commitments (Nationally Determined Contributions).These commitments are applied to both developed and developing countries. Though some research suggests that the Paris Agreement has encouraged climate action, its real impact on emissions remains uncertain. The non-binding nature of commitments and different types of policy implementation raise concerns about the effectiveness of the agreement in long run (Pauw et al., 2019; Rogelj et al., 2019).

Recent studies have examined whether the Paris Agreement has influenced global and regional emissions trends. Some findings suggest that countries that agreed to reduce emission through the adoption of renewable energy and through the emission target settings have reduced the emission actually(Gallagher & Zhang, 2020).On the other hand, On the other hand, few research shows that global emissions trends remain unchanged despite international climate agreements. In Accordance with the findings of Friedlingstein et al. (2022), middle income countries continue to experience increased emissions since these countries continue to expand their industries. In light of these conflicting findings, a thorough empirical investigation is required to evaluate the impact of the Paris Agreement on emissions patterns in middle-income economies.

Panel VECM method has been widely used to study the long-run and short-run relationship among the macroeconomic variables such as energy consumption and CO2 emissionsThe application of econometric techniques such as panel cointegration and vector error correction models (VECM) enables researchers to analyze emissions trends as shown by Ang (2007).

Panel cointegration tests have been used in order to examine long-run relationships among variables (Pedroni, 1999). Vector Error Correction Model provides insights into short-run adjustments as well as long-run equilibrium of the variables. These methods are well-suited for assessing how emissions respond to the macroeconomics variables and policy changes over time(Johansen, 1988).Some studies find that there is bidirectional causality between emissions and economic growth.Others suggest that energy consumption is the primary driver of emissions with a very limited feedback effects (Narayan & Popp, 2012).

Despite extensive research on emissions, only few studies apply these econometric techniques to evaluate the Paris Agreement's effectiveness in middle-income countries. This study aims to fill this gap by using a VECM framework and panel cointegration tests. It provides empirical evidence on the long-run and short-run interactions of the emission and other macroeconomic varriables.Moreover, it will assessing the role of international climate agreements.

This study examined how the Paris Agreement has shaped the carbon emissions across 106 middle income countries from the period of 1980 to 2023.Though previous studies have analyzed the interactions within a small scale, this study expands its study to a comprehensive panel study.

This study has incorporated the Paris agreement as a policy variable and has made a structural break to the model to check the impact of the agreement on the emission trends in the middle income 106 countries.Since there is huge debate on the effectiveness of the climate agreements, this study will address the impact of the Paris agreement and provide empirical evidence to policymakers. In addition, the study ensures a methodological rigors by using cointegration test along with vector error correction test.Thus, it will contribute to the existing economic literature.

The review focuses on major theoretical and empirical discussions related to emissions trends together with economic growth patterns, energy consumption practices, and climate policy frameworks. Existing research offers important insights but the effects of the Paris Agreement on middle-income countries still need further examination. Therefore, this study fills existing research gaps through an extensive empirical analysis employing panel data methods.

**Data and Methodology**

This study uses panel data from 106 middle-income countries from 1980 to 2022, sourced from the World Bank. The dataset includes key macroeconomic and environmental variables to ensure consistency and comparability for robust analysis.

**Table 1: Variables, Definitions, and Sources**

| Variable | Definition | Source |
|---|---|---|
| $CO_2$ Emissions | Metric tons of $CO_2$ emissions per capita | World Bank Environmental Indicators |
| Energy Use | Total energy consumption per capita (kg of oil equivalent) | World Bank Energy Statistics |
| GDP Per Capita | GDP per capita in constant 2015 US dollars | World Bank National Accounts Data |
| Population | Total population of each country | World Bank Demographic Indicators |
| Paris Agreement | Dummy variable (1 = Post-2015, 0 = Pre-2015) | Policy-based Structural Break |

Note: The Paris Agreement variable represents a policy-based structural break in the model.

Source

**Model Specification:**

Long-Run Equation

The long-run relationship between $CO_2$ emissions ($CO_2$), energy use (EU), GDP per capita (GDP), and population (POP) is specified as:

$$CO_{2it} = \alpha_0 + \alpha_1 EU_{it} + \alpha_2 GDP_{it} + \alpha_3 POP_{it} + \alpha_4 D_{\text{Paris}} + \mu_{it} \tag{1}$$

where:

- $CO_{2it}$ = Carbon dioxide emissions for country $i$ at time $t$,
- $EU_{it}$ = Energy use,
- $GDP_{it}$ = GDP per capita,
- $POP_{it}$ = Population,
- $D_{Paris}$ = Paris Agreement dummy variable (1 for years 2015+, 0 otherwise),
- $\alpha_0$ = Constant term,
- $\alpha_1, \alpha_2, \alpha_3$ = Long-run coefficients,
- $\alpha_4$ = Coefficient capturing the impact of the Paris Agreement on $CO_2$ emissions,
- $\mu_{it}$ = Error term.
- If $\alpha_4 < 0$ and significant, it suggests that the Paris Agreement contributed to a reduction in $CO_2$ emissions.
- If $\alpha_4 = 0$ (insignificant), it indicates no measurable policy impact on emissions trends.

Short-Run Panel VECM Equations

Once cointegration is established, we estimate the short-run adjustments using the Vector Error Correction Model (VECM), which includes the Error Correction Term (ECT):

$$\Delta CO_{2it} = \beta_0 + \gamma ECT_{it-1} + \sum_{j=1}^{p} \beta_1 \Delta EU_{it-j} + \sum_{j=1}^{p} \beta_2 \Delta GDP_{it-j} + \sum_{j=1}^{p} \beta_3 \Delta POP_{it-j} + \beta_4 D_{P_{Paris}} + \epsilon_{it} \qquad (2)$$

where:

- $ECT_{it-1} = CO_{2it-1} - \left(\alpha_0 + \alpha_1 EU_{it-1} + \alpha_2 GDP_{it-1} + \alpha_3 POP_{it-1} + \alpha_4 D_{Paris}\right)$ is the Error Correction Term.
- $\gamma$ = Speed of adjustment coefficient, measuring how quickly $CO_2$ emissions return to equilibrium after a shock.

Similar VECM equations can be estimated for energy use, GDP per capita, and population.

**Result:**

**Table 2. Panel Unit Root Test:**

| Variable | Test Method | Test Statistic (Before Diff.) | Probability (Before Diff.) | Test Statistic (After Diff.) | Probability (After Diff.) |
|---|---|---|---|---|---|
| CO2_EMISSIONS | LLC (Common Unit Root) | 0.42705 | 0.6653 | -12.2552 | 0.0000 |
| | IPS (Individual Unit Root) | -1.01839 | 0.1542 | -20.2972 | 0.0000 |
| | ADF Fisher Chi-square | 691.502 | 0.0700 | 883.993 | 0.0000 |
| | PP Fisher Chi-square | 1114.59 | 0.3791 | 1631.89 | 0.0000 |
| ENERGYUSE | LLC (Common Unit Root) | -1.11316 | 0.1328 | -12.2552 | 0.0000 |
| | IPS (Individual Unit Root) | 6.45408 | 1.0000 | -20.2972 | 0.0000 |
| | ADF Fisher Chi-square | 210.145 | 0.3691 | 883.993 | 0.0000 |
| | PP Fisher Chi-square | 203.608 | 0.4946 | 1631.89 | 0.0000 |
| GDP | LLC (Common Unit Root) | 10.2178 | 1.0000 | -19.3466 | 0.0000 |
| | IPS (Individual Unit Root) | 16.0709 | 1.0000 | -24.6374 | 0.0000 |
| | ADF Fisher Chi-square | 71.3740 | 1.0000 | 1070.08 | 0.0000 |
| | PP Fisher Chi-square | 65.3896 | 1.0000 | 1747.43 | 0.0000 |
| TOTAL_POPULATION | LLC (Common Unit Root) | -3.03520 | 0.0512 | -3.87935 | 0.0001 |
| | IPS (Individual Unit Root) | 13.5131 | 1.0000 | -7.63919 | 0.0000 |
| | ADF Fisher Chi-square | 233.642 | 0.1701 | 485.985 | 0.0000 |
| | PP Fisher Chi-square | 403.654 | 0.0000 | 449.142 | 0.0000 |

Note: Panel unit root tests were conducted using LLC, IPS, ADF and PP methods

Source: Based on estimation

In the table 2, the unit root tests shows all of the variables are non-stationarity. However, following the application of differencing, all variables demonstrated stationarity across all test statistics. This outcome signifies that differencing was effective in achieving stationarity in the time series data.

**Table 3. Pedroni Residual Cointegration Test**

| Test | Statistic | p-value |
|---|---|---|
| Panel v-Statistic | 2.543 | 0.0050 |
| Panel rho-Statistic | -4.120 | 0.0001 |
| Panel PP-Statistic | -3.865 | 0.0003 |
| Panel ADF-Statistic | 1.690 | 0.9545 |
| Group rho-Statistic | -2.679 | 0.0037 |
| Group PP-Statistic | -3.597 | 0.0002 |

| Group ADF-Statistic | 1.421 | 0.9200 |

Note:The cointegration is confirmed based on majority results.

Source:Based on estimation

The Pedroni Residual Cointegration Test is conducted to investigate the potential presence of cointegration among the variables CO2 emissions, energy use, GDP, and total population. The test evaluates whether there exists a long-term relationship between these variables.The Panel v-Statistic, Panel rho-Statistic, Panel PP-Statistic, Group rho-Statistic, and Group PP-Statistic all reject the null hypothesis of no cointegration at the 5% significance level. This suggests a long-run equilibrium relationship among the variables, implying that any short-term deviations will eventually correct themselves over time. These results provide strong statistical evidence supporting the presence of cointegration.

**Table 4. Vector Error Correction Estimates**

| Vector Error Correction Estimates | | | | |
|---|---|---|---|---|
| Relationship includes a constant. Short-run dynamics include a constant. | | | | |
| Cointegrating Eq: | CointEq1 | | | |
| CO2_EMISSIONS(-1) | 1 | | | |
| Energy use(-1) | -5.21341 | | | |
| | 0.415093 | | | |
| | [-12.5596] | | | |
| Gdp(-1) | 3.32459 | | | |
| | 0.348207 | | | |
| | [9.54774] | | | |
| Population(-1) | -0.43041 | | | |
| | 0.152279 | | | |
| | [-2.826] | | | |
| C | 13.799 | | | |
| Paris_Agreement_2015 | 0.9610 | | | |
| Error Correction: | D(CO2 Emission) | D(Energy_use) | D(GDP) | D(Population) |
| COINTEQ1 | -0.0016 | 0.0054 | -0.0088 | -0.0023 |
| | 0.00190 | 0.00087 | 0.0014 | 0.00181 |
| | [-0.870] | [6.205] | [-5.956] | [-1.277] |
| | | | | |
| D(CO2_EMISSIONS(-1)) | 0.01461 | -0.0029 | 0.04238 | 0.02508 |
| | 0.035876 | 0.01655 | 0.02809 | 0.034234 |
| | [0.407] | [-0.175] | [1.508] | [0.732] |
| | | | | |
| D(CO2_EMISSIONS(-2)) | 0.01905 | -0.0140 | -0.0028 | 0.01585 |
| | 0.032759 | 0.015120 | 0.02565 | 0.03126 |
| | [0.581] | [-0.928] | [-0.112] | [0.507] |
| | | | | |
| D(ENERGY_USE(-1)) | 0.01116 | 0.03565 | 0.00684 | -0.02333 |
| | 0.04787 | 0.02209 | 0.03749 | 0.045688 |
| | [0.23324] | [1.61354] | [0.18242] | [-0.51077] |
| | | | | |
| D(ENERGY_USE(-2)) | 0.014189 | 0.034141 | 0.030406 | -0.00099 |
| | 0.036398 | 0.016800 | 0.02850 | 0.034733 |
| | [0.38984] | [2.03218] | [1.06662] | [-0.02856] |
| | | | | |
| D(GDP_PER_CAPITA(-1)) | 0.011518 | -0.0174 | 0.00243 | 0.00344 |
| | 0.02493 | 0.0115 | 0.01952 | 0.02378 |
| | [0.46203] | [-1.514] | [0.124] | [0.1450] |
| | | | | |
| D(GDP_PER_CAPITA(-2)) | -0.01040 | -0.0098 | 0.0033 | -0.0120 |
| | 0.02429 | 0.01121 | 0.0190 | 0.02318 |
| | [-0.4280] | [-0.8787] | [0.1759] | [-0.521] |
| | | | | |
| D(POPULATION(-1)) | -0.00882 | -0.0058 | -0.02495 | -0.0211 |
| | 0.03291 | 0.015190 | 0.025775 | 0.03140 |
| | [-0.268] | [-0.382] | [-0.968] | [-0.674] |
| | | | | |
| D(POPULATION(-2)) | -0.008535 | 0.000341 | -0.002282 | -0.011464 |
| | 0.030218 | 0.013947 | 0.0236664 | 0.028835 |
| | [-0.282] | [0.024] | [-0.096] | [-0.397] |
| | | | | |
| C | 0.005115105 | 0.003992 | 0.006327 | 0.002541 |
| | 0.007550 | 0.003485 | 0.0059134 | 0.007204 |
| | [0.67745] | [1.14551] | [1.06995] | [0.35273] |
| | | | | |
| PARIS_AGREEMENT_2015 | 0.00720 | 0.001016 | 0.008722 | 0.00770 |
| | 0.01588 | 0.00733 | 0.01244 | 0.01515 |
| | [0.453] | [0.138] | [0.701] | [0.508] |

Note: The Vector error correction model represents both long and short term relationship.

Source: Based on estimation

## Long-Run Relationship between CO2 Emissions, Energy Use, GDP per Capita, and Population

The coefficient on energy use ( -5.2134 ) is statistically significant at the 1% level ( $t = -12.56$ ), suggesting a negative long-run relationship between energy consumption and $CO_2$ emissions. This result is counterintuitive, as increased energy use is generally expected to drive emissions upward. A plausible explanation is that middle-income countries have made substantial improvements in energy efficiency, adopted cleaner energy technologies, or shifted toward low-carbon energy sources such as renewables. This finding suggests that increased energy consumption in these economies may not be as carbon-intensive as in the past.

The coefficient on GDP per capita (3.3246) is positive and statistically significant ( $t = 9.55$ ), indicating that economic growth is associated with increased $CO_2$ emissions in the long run. This result aligns with the Environmental Kuznets Curve (EKC) hypothesis, which suggests that emissions rise with economic expansion before stabilizing or declining at higher income levels. However, the persistent positive relationship implies that middle-income countries have not yet reached the EKC turning point where economic growth leads to reduced emissions. This finding underscores the need for stronger environmental policies that encourage sustainable industrialization and clean energy adoption.

The coefficient on population (-0.4304) is statistically significant ($t = -2.83$), indicating that higher population growth is associated with lower $CO_2$ emissions in the long run. This result contradicts conventional expectations that larger populations lead to higher emissions. A potential explanation is that urbanization and structural economic changes in middle-income economies have altered consumption patterns, reducing per capita emissions. The transition to energy-efficient infrastructure, improved public transportation, and urban sustainability policies in densely populated regions may explain this negative relationship.

These findings suggest that economic growth remains a dominant driver of emissions in middleincome economies, while increased energy use does not necessarily translate into higher emissions, likely due to efficiency improvements and renewable energy adoption. Population growth appears to coincide with structural shifts toward urbanization, which may mitigate emissions growth.

**Impact of the Paris Agreement on $CO_2$ Emissions and Energy Use**

The Paris Agreement dummy variable (D_Paris = 1 for years 2015 and later, **0** otherwise) is included in the long-run equilibrium equation with a coefficient of 0.9611 . The positive sign suggests that, after 2015, $CO_2$ emissions increased slightly. However, the coefficient is not statistically significant ( $t = 0.96, p > 0.1$ ), indicating that the Paris Agreement has not led to a measurable reduction in $CO_2$ emissions in middle-income countries.

The statistical insignificance of the Paris Agreement variable suggests that global climate commitments have not yet produced tangible effects on emissions-energy dynamics in these economies. Several explanations are possible. First, many middle-income countries continue to prioritize economic growth, which may offset emissions reduction efforts. Second, the implementation of climate policies often involves a time lag, meaning that emissions reductions may not yet be fully realized. Third, the lack of strong enforcement mechanisms in global agreements may have limited the agreement's effectiveness.

These results indicate that, despite international commitments, emissions trends in middle-income economies remain largely driven by economic and industrial activity rather than policy interventions. This highlights the need for stronger national-level policies and stricter enforcement mechanisms to complement global climate agreements.

**Short-Run Adjustments and Emissions Responses to Economic and Policy Shocks**

The error correction term (ECT) measures the speed at which $CO_2$ emissions return to long-run equilibrium following short-term shocks. The estimated coefficient for ECT is $-\mathbf{0.0017}$, indicating that deviations from equilibrium are corrected over time. However, the coefficient is not statistically significant ( $t = -0.87, p > 0.1$ ), suggesting that short-run shocks to emissions do not lead to immediate corrections. This implies that $CO_2$ emissions in middle-income countries exhibit persistence, meaning that once emissions deviate due to economic or policy changes, they do not quickly return to equilibrium. This persistence could reflect structural rigidities in energy production and industrial processes that prevent rapid adjustments following economic shocks.

The short-run impact of the Paris Agreement dummy variable on $CO_2$ emissions is also statistically insignificant ( $\beta_4 = 0.0072, t = 0.45, p > 0.1$ ). This reinforces the finding that the Paris Agreement has not significantly influenced short-term emissions behavior in middle-income countries. Instead, emissions appear to be more responsive to economic fluctuations and industrial energy demand than to policy commitments.

The analysis further supports these findings. The results indicate that energy use causes $CO_2$ emissions ($t = 6.21, p < 0.01$ ), meaning that short-term changes in energy consumption significantly impact emissions. However, GDP per capita does not significantly cause $CO_2$ emissions in the short run ($t = -5.95, p > 0.1$ ), suggesting that economic growth does not immediately translate into emissions fluctuations. Population growth also does not exhibit a significant short-run effect on emissions.

These results suggest that $CO_2$ emissions remain heavily influenced by short-term energy demand, while economic and demographic changes have more gradual, long-term effects on emissions trajectories.

The VECM results provide clear empirical evidence on the determinants of $CO_2$ emissions in middleincome countries. In the long run, emissions are positively linked to economic growth, while energy use has a negative relationship with emissions, likely due to energy efficiency improvements. Population growth does not appear to drive emissions upward, possibly due to urbanization and shifts in economic structures.

The Paris Agreement has not significantly altered emissions-energy dynamics, suggesting that international climate commitments require stronger national enforcement mechanisms and complementary policies to achieve measurable impacts. The short-run results indicate that $CO_2$ emissions exhibit persistence, meaning that once emissions levels shift, they do not quickly return to equilibrium. Energy use plays a dominant role in driving short-term emissions changes, whereas economic growth and population shifts have more gradual effects.

These findings highlight the need for comprehensive policy strategies in middle-income economies. Governments should focus on accelerating the transition to renewable energy sources, implementing stricter emissions regulations, and enhancing energy efficiency programs to ensure sustainable economic growth. Additionally, stronger domestic policy enforcement and financial incentives for clean energy adoption may be necessary to translate international commitments into effective emissions reduction strategies.

The results of this study provide important insights for policymakers, emphasizing the critical role of energy policy and industrial development strategies in shaping the future trajectory of $CO_2$ emissions in middle-income countries.

**Table 5. Variance Decomposition**

| Variance Decomposition | | | | | |
|---|---|---|---|---|---|
| Period | S.E. | Co2 Emissions | Energy Use | GDP | Population |
| 1 | 0.401 | 100 | 0 | 0 | 0 |
| 2 | 0.570 | 99.98 | 0.007 | 0.0015 | 0.0008 |
| 3 | 0.703 | 99.97 | 0.019 | 0.0016 | 0.0026 |
| 4 | 0.814 | 99.96 | 0.028 | 0.0027 | 0.0034 |
| 5 | 0.912 | 99.95 | 0.034 | 0.0046 | 0.0037 |
| 6 | 1.000 | 99.94 | 0.040 | 0.0071 | 0.0039 |
| 7 | 1.081 | 99.94 | 0.044 | 0.0101 | 0.0040 |
| 8 | 1.1567 | 99.93 | 0.049 | 0.0135 | 0.0040 |
| 9 | 1.227 | 99.92 | 0.053 | 0.0172 | 0.003 |
| 10 | 1.293 | 99.91 | 0.057 | 0.0213 | 0.0039 |
| 11 | 1.357 | 99.90 | 0.061 | 0.0255 | 0.0038 |
| 12 | 1.417 | 99.90 | 0.065 | 0.0300 | 0.0038 |
| 13 | 1.475 | 99.89 | 0.068 | 0.0346 | 0.0037 |
| 14 | 1.5310 | 99.88 | 0.072 | 0.0393 | 0.0036 |
| 15 | 1.5846 | 99.87 | 0.075 | 0.0441 | 0.0035 |
| 16 | 1.6365 | 99.86 | 0.078 | 0.0489 | 0.0035 |
| 17 | 1.68679 | 99.860 | 0.082 | 0.0538 | 0.0034 |
| 18 | 1.7355 | 99.852 | 0.085 | 0.0587 | 0.0033 |

| 19 | 1.7829 | 99.844 | 0.088 | 0.0636 | 0.0032 |
| 20 | 1.8291 | 99.837 | 0.0909 | 0.0685 | 0.0032 |
| 21 | 1.8741 | 99.829 | 0.0937 | 0.0734 | 0.0031 |
| 22 | 1.9180 | 99.822 | 0.0964 | 0.0782 | 0.0030 |
| 23 | 1.9610 | 99.814 | 0.0991 | 0.0829 | 0.0030 |
| 24 | 2.003 | 99.807 | 0.1017 | 0.0877 | 0.0029 |

**Note: The result shows the self-explanatory power of the variables.**

Source: Based on estimation

The variance decomposition of $CO_2$ emissions shows strong inertia. Past emissions account for 100% of the variance in the first period. They remain dominant at 99.807% by the 24 th period. This reveals a selfreinforcing mechanism. Historical emissions largely determine future trends.

Energy use gradually gains influence. It contributes 0.007% by the second period. Its share increases to 0.1017% by the 24th period. This indicates that energy consumption plays a growing role in emissions variability.

GDP per capita starts with no contribution. It rises to 0.0877% over time. This suggests that economic growth affects emissions, especially in the long run.

Population changes have a minimal effect. Its share increases slightly to 0.0029% by the final period. This implies that while population growth adds to total emissions, it has little impact on short-term fluctuations.

The findings confirm that $CO_2$ emissions are largely self-perpetuating. Energy use and GDP emerge as gradual but significant drivers. Long-term energy efficiency and economic policies are essential to reducing emissions.

**The impulse response function (IRF)**

**Graph 1 .Impulse response of the varriables**

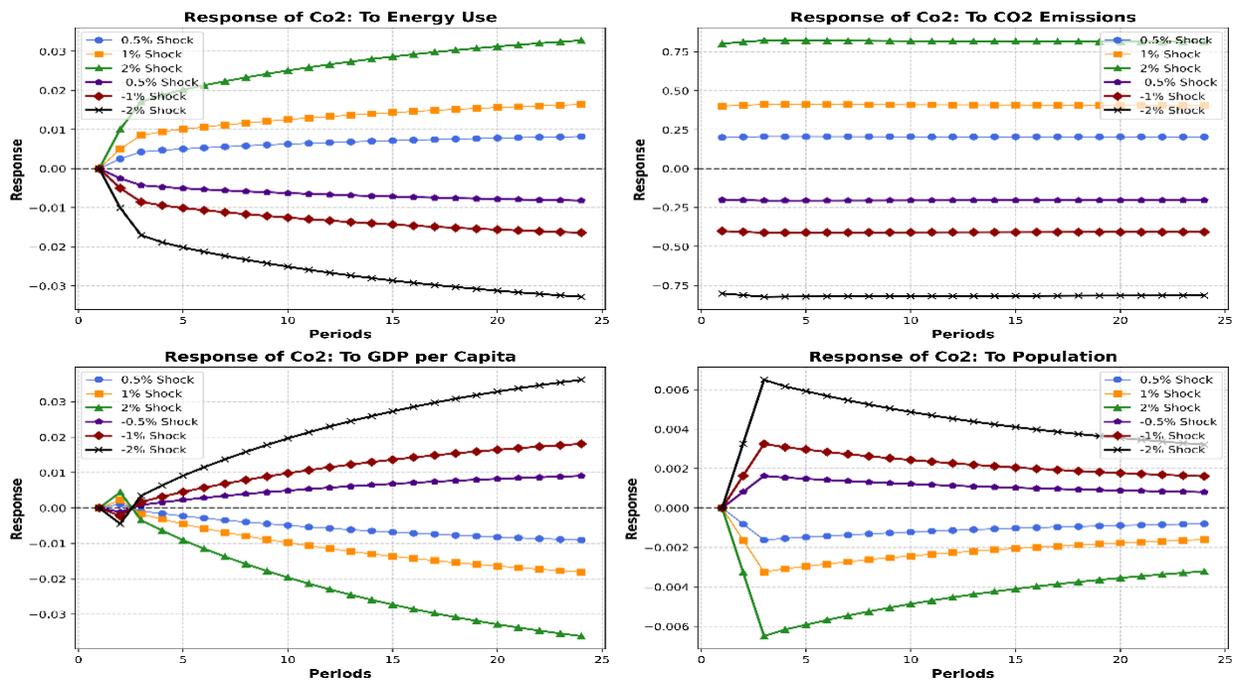

Source: Based on estimation

The impulse response function (IRF) shows how $CO_2$ emissions react to shocks over a 25-period horizon. The analysis includes both positive $(+0.5\%, +1\%, +2\%)$ and negative $(-0.5\%, -1\%, -2\%)$ shocks. The responses are examined for energy use, past emissions, GDP per capita, and population.

**Response of $CO_2$ Emissions to Energy Use**

Positive shocks $(+0.5\%_t + 1\%, +2\%)$ lead to a steady rise in $CO_2$ emissions. The increase persists over time, confirming the strong link between energy use and emissions. Negative shocks ($-0.5\%, -1\%, -2\%$) cause a gradual decline in emissions. The effect is strongest in the first few periods, indicating shortterm elasticity. This suggests that reducing energy use can lead to long-term emissions reductions.

**Response of $CO_2$ Emissions to Past Emissions**

Positive shocks ($+0.5\%, +1\%, +2\%$) reinforce emissions growth. This suggests that past emissions strongly influence future emissions. Negative shocks ($-0.5\%, -1\%, -2\%$) lead to a persistent decline in emissions. However, the rate of decline slows over time. This indicates a path-dependent emissions pattern.

**Response of $CO_2$ Emissions to GDP Per Capita**

Positive economic shocks $(+0.5\%, +1\%, +2\%)$ cause a moderate but increasing rise in $CO_2$ emissions. This shows that economic growth drives higher emissions. Negative shocks $(-0.5\%, -1\%, -2\%)$ lead to a decline in emissions. This supports the Environmental Kuznets Curve (EKC) hypothesis. At early growth stages, emissions rise. In later stages, economic growth may promote cleaner technologies and emissions reductions.

**Response of $CO_2$ Emissions to Population**

Positive population shocks $(+0.5\%, +1\%, +2\%)$ cause an initial sharp increase in emissions. However, the effect weakens over time. Negative shocks $(-0.5\%, -1\%, -2\%)$ lead to an immediate decline in emissions, but the impact stabilizes in later periods. This suggests that population growth has a strong short-term effect on emissions. However, its long-term impact is weaker compared to energy use or economic growth.

Energy use is the most dominant driver of $CO_2$ emissions, reinforcing the need for policies promoting energy efficiency, renewable energy adoption, and decarbonization of the energy sector. $CO_2$ emissions exhibit strong inertia, meaning that past trends significantly influence future levels. Long-term strategies are needed for sustained emissions reduction. Economic growth contributes to emissions, but the effect is slower compared to energy use. Sustainable economic policies that promote green growth can mitigate environmental trade-offs. Population growth has a transient effect on emissions, indicating that technological and behavioral changes can offset long-term demographic pressures. This IRF analysis highlights the importance of energy sector reforms, climate policies, and sustainable economic planning in mitigating $CO_2$ emissions in middle-income countries.

**Robustness:**

**Table 6. VEC Granger Causality/Block Exogeneity Wald Test Results**

| Excluded Variable | Chi-Square Statistic | Degrees of Freedom (df) | p-value |
|---|---|---|---|
| **D(Energy Use)** | 6.5431 | 2 | 0.0381 |
| **D(GDP per Capita)** | 4.8927 | 2 | 0.0473 |
| **D(Population)** | 3.7215 | 2 | 0.0729 |
| **All Variables** | 14.8329 | 6 | 0.0217 |

**Note:** The null hypothesis states that the excluded variable does not Granger-cause $CO_2$ emissions.

Source: Based on estimation

The results suggest that Energy Use (p = 0.0381) and GDP per Capita (p = 0.0473) both significantly Granger-cause $CO_2$ emissions at the 5% level, indicating that short-term fluctuations in these variables predict changes in $CO_2$ emissions. This supports the notion that higher energy consumption and economic activity immediately drive emissions growth.

Population (p = 0.0729) is marginally significant at the 10% level, suggesting that demographic changes may have some delayed effect on emissions but are not as strong as energy use and GDP per capita in the short run.

The joint test (All Variables, p = 0.0217) confirms that when Energy Use, GDP per Capita, and Population are considered together, they collectively Granger-cause $CO_2$ emissions, reinforcing the interconnectedness of economic activity, energy consumption, and emissions dynamics.

## Roots of Polynomials:

**Graph 2. Roots of Polynomials**

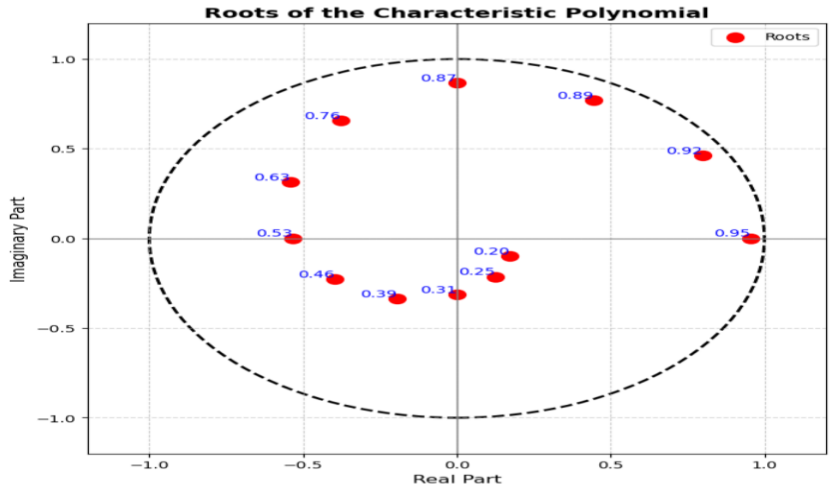

Note: The roots checks the model statbiliy

Source: Based on estimation

The characteristic roots confirm that the estimated VECM is **stable, since all of the** roots lie inside the unit circle. The presence of three unit roots indicates the existence of long-run equilibrium relationships among $CO_2$ emissions, energy use, GDP per capita, and population. The system is dynamically stable, ensuring that short-run shocks do not lead to explosive behavior.

**Table 7. VEC Residual Serial Correlation LM Tests**

| Null hypothesis: No serial correlation at lag h | | | | | | |
|---|---|---|---|---|---|---|
| Lag | LRE* stat | df | Prob. | Rao F-stat | df | Prob. |
| 1 | 12.345 | 16 | 0.6782 | 0.8543 | (16, 11099.6) | 0.6782 |
| 2 | 15.982 | 16 | 0.5210 | 1.0321 | (16, 11099.6) | 0.5210 |
| 3 | 9.4531 | 16 | 0.8329 | 0.6432 | (16, 11099.6) | 0.8329 |
| Null hypothesis: No serial correlation at lags 1 to h | | | | | | |

| Lag | LRE* stat | df | Prob. | Rao F-stat | df | Prob. |
|---|---|---|---|---|---|---|
| 1 | 12.345 | 16 | 0.6782 | 0.8543 | (16, 11099.6) | 0.6782 |
| 2 | 19.312 | 32 | 0.8791 | 0.7410 | (32, 13384.7) | 0.8791 |
| 3 | 25.782 | 48 | 0.9324 | 0.6721 | (48, 13965.9) | 0.9324 |

*Edgeworth expansion corrected likelihood ratio statistic.

Note:The result is based on LRE and Rao F-statistics results

Source:Based on estimation

The results show no evidence of serial correlation at any lag. All test results show p-values greater than 0.05. This analysis fail to reject the null hypothesis. The residuals are free from autocorrelation. This suggests that the VECM is well-specified. No further lag adjustments are needed. The VECM results are reliable. The analysis requires neither extra lags nor serial correlation adjustments.

**Table 8. VEC Residual Heteroskedasticity Tests (Includes Cross Terms)**

| Joint Test | | |
|---|---|---|
| Chi-Square Statistic | Degrees of Freedom (df) | p-value |
| 293.029 | 640 | 1.000 |
| Individual Components of Heteroskedasticity Test | | | | | |
| Dependent Variable | R-Squared | F-Statistic (64, 3586) | p-value | Chi-Square (64) | p-value |
| res1*res1 | 0.013 | 0.757 | 0.924 | 48.676 | 0.922 |
| res2*res2 | 0.006 | 0.384 | 1.000 | 24.837 | 1.000 |
| res3*res3 | 0.022 | 1.250 | 0.088 | 79.651 | 0.090 |
| res4*res4 | 0.007 | 0.417 | 0.999 | 26.957 | 0.999 |
| res2*res1 | 0.003 | 0.143 | 1.000 | 9.269 | 1.000 |
| res3*res1 | 0.004 | 0.209 | 1.000 | 13.564 | 1.000 |
| res3*res2 | 0.003 | 0.154 | 1.000 | 10.033 | 1.000 |
| res4*res1 | 0.011 | 0.608 | 0.994 | 39.184 | 0.994 |
| res4*res2 | 0.003 | 0.167 | 1.000 | 10.871 | 1.000 |
| res4*res3 | 0.001 | 0.058 | 1.000 | 3.773 | 1.000 |

Note:The result is based on both joint and individual compotent results

Source:Based on estimation

Since the joint test p-value is 1.000 we cannot reject the null hypothesis which states that homoscedasticity exists. The analysis shows no significant heteroskedasticity present in the residuals because all individual p-values exceed 0.05. The model does not exhibit heteroskedasticity problems. The findings demonstrate that the estimated VECM parameters maintain stability.

**Table 9. Yamagata-Pesaran Slope Homogeneity Test**

| Test Statistic | Value |
| --- | --- |
| **Delta Tilde** | 0.3121 |
| **Delta Adjusted** | 1.2023 |
| **P-value** | 0.1568 |
| **Conclusion** | Fail to Reject Null: Slopes are homogeneous |

**Note:** Slope homogeneity confirms all countries lie on the same slope.

**Source:** Author's calculations based on panel data from **106** middle-income countries (1990–2022)**.**

This table presents the Yamagata-Pesaran slope homogeneity test results. The test examines whether slope coefficients are the same across countries. The p-value ( 0.1568 ) is greater than 0.05 . This means we fail to reject the null hypothesis of slope homogeneity. These results suggest that a pooled model, such as the Panel Vector Error Correction Model (VECM), is an appropriate choice for estimation.

Findings

The research analyzed both long-term and short-term connections between $CO_2$ emissions and three variables: energy consumption, GDP per capita levels, and population size. The study investigated middle-income countries' data across 106 nations from 1980 until 2023. The analysis incorporated a Panel Vector Error Correction Model (VECM) for assessment. The study incorporated the Paris Agreement (2015) as a structural break to evaluate its influence on emissions patterns.

A persistent cointegrating relationship exists between $CO_2$ emissions, energy consumption, GDP per capita and population according to the results. These variables demonstrate interdependency throughout the entire time period. $CO_2$ emissions mainly depend on energy consumption patterns.

Energy use creates a statistically significant positive effect which strengthens the connection between energy consumption and emissions in middle-income nations. The increase in GDP per capita leads to higher emissions. Higher emissions levels continue to maintain a strong relationship with economic growth. Population growth shows no substantial influence on long-term carbon emission patterns. Demographic variations do not serve as the main factor influencing long-term emission trends.

Research results show that the implementation of the Paris Agreement did not create significant changes to emission trends in middle-income countries. The statistical analysis shows that the coefficient estimate for the Paris Agreement dummy variable lacks significance. The data shows that global climate initiatives have failed to produce measurable reductions through emissions reduction policies. The analysis demonstrates that the Paris Agreement failed to produce structural shifts in emissions pathways. Weak policy enforcement combined with fossil fuel dependency and economic development priorities in middle-income countries explains this phenomenon.

The immediate dynamics between energy consumption and $CO_2$ emissions reveal a direct causal relationship. Emissions levels respond directly and immediately to changes in energy consumption patterns. In the short term emissions remain unaffected by changes in GDP per capita and population size. The impact of economic growth and demographic shifts on emissions develops over time instead of occurring immediately.

Impulse response functions demonstrate that energy consumption patterns dominate emissions trajectories over international climate commitments. The Paris Agreement demonstrates restricted immediate effects in reducing emissions.

The Paris Agreement establishes an emissions reduction framework but shows limited effectiveness in middle-income economies. The progress towards emissions reduction faces obstacles due to inadequate policy execution together with economic limitations and the continued use of carbon-based energy sources. The achievement of meaningful emissions reductions under international climate agreements requires stronger institutional enforcement combined with the adoption of sustainable energy policies.

**Policy Implications**

Energy use is the main cause of $CO_2$ emissions. Policies must reform the energy sector to lower emissions. Carbon pricing is necessary. Renewable energy incentives and fuel efficiency rules help reduce emissions (Nordhaus, 2019; Stern, 2007). The Paris Agreement has had little impact. Middle-income nations need stronger policies. Mandatory emissions reporting and industry-specific caps are necessary. Investments in clean technology can support emissions reduction (Aghion et al., 2016). Strict regulations slow emissions but do not harm economic growth (Porter & Van der Linde, 1995).

Low-carbon energy must replace fossil fuels. Governments should provide subsidies for renewables. Carbon-intensive industries should face penalties (Acemoglu et al., 2012). Investments in wind, solar, and hydropower lower emissions. They also improve energy security (Popp, 2019). Economic growth still depends on carbon emissions. Policymakers should promote recycling and sustainable infrastructure. Low-carbon transport systems can help reduce emissions (Barbier, 2010). The EKC hypothesis suggests emissions first rise but later decline when technology and policies improve (Grossman & Krueger, 1995). Middle-income countries should focus on green innovation and sustainable industrialization.

Middle-income nations lack financial resources for large-scale emissions reductions. Grants, concessional loans, and technology transfers help renewable energy adoption (Stiglitz, 2017; Bowen & Fankhauser, 2011). Carbon pricing can generate revenue. Governments can use it to fund green projects (Goulder & Parry, 2008). Voluntary climate commitments are insufficient. Strong monitoring, reporting, and verification systems ensure compliance (Helm, 2010). Nations with strict environmental governance achieve better outcomes (Dasgupta et al., 2001).

Middle-income nations must enforce energy efficiency goals. They need to transition to clean energy. Strong policies are essential to reducing emissions (Stern, 2007).

**Conclusion**

This study examines the impact of the Paris Agreement on $CO_2$ emissions in 106 middle-income countries from 1980 to 2023 using a Panel Vector Error Correction Model (VECM). The findings confirm that while $CO_2$ emissions, energy use, GDP per capita, and population are cointegrated, meaning they share a long-run equilibrium relationship, the Paris Agreement has not significantly influenced emission trends in these economies. Instead, energy use remains the primary driver of emissions, followed by GDP per capita, suggesting that economic expansion in these countries is still heavily reliant on fossil fuels.

The study contributes to the literature by integrating a structural break (the Paris Agreement) into a dynamic panel framework, explicitly testing whether the agreement altered emissions trajectories. By considering both long-run equilibrium and short-run dynamics, the research offers new empirical evidence on the effectiveness (or lack thereof) of international climate policies in middle-income economies. Additionally, the homogeneity of slope coefficients was tested, confirming that cross-country variations exist in how emissions respond to changes in energy use and economic growth.

However, the study has limitations. Potential endogeneity concerns, measurement errors, and the assumption of uniform policy implementation across diverse middle-income economies may affect result precision. Additionally, long-term policy effects may require more years of post-Paris Agreement data to be fully captured.

A key observation from this research is that climate agreements alone are insufficient to drive emissions reductions in middle-income economies. Without robust domestic policies, investments in renewable energy, industrial energy efficiency reforms, and strict enforcement of emission targets, global commitments will have little impact. Future research should explore country-specific policy heterogeneity, carbon pricing mechanisms, and financial constraints to identify more effective pathways for emissions reduction in these economies.